\documentclass[preprint,aps,amsmath,amssymb,floatfix]{revtex4}
\usepackage{graphicx}
\usepackage{dcolumn}
\usepackage{bm}
\usepackage{subfigure}
\usepackage{epsfig}

\renewcommand{\i}{{\rm i}}
\usepackage{color}
\usepackage{txfonts}

\begin{document}


\title{Classically Isospinning Hopf Solitons}

\author{Richard A.~Battye}
\email{richard.battye@manchester.ac.uk}
\author{Mareike Haberichter}%
 \email{mkh@jb.man.ac.uk}
\affiliation{%
Jodrell Bank Centre for Astrophysics, University of Manchester M13 9PL, U.K.\\
}%

\date{\today}
        
\begin{abstract}

We perform full three-dimensional numerical relaxations  of isospinning Hopf solitons with Hopf charge up to 8 in the Skyrme-Faddeev model with mass terms included. We explicitly allow the soliton solution to deform and to break the symmetries of the static configuration. It turns out that the model with its rich spectrum of soliton solutions, often of similar energy, allows for transmutations, formation of new solution types and the rearrangement of the spectrum of minimal-energy solitons in a given topological sector when isospin is added. We observe that the shape of isospinning Hopf solitons can differ qualitatively from that of the static solution. In particular the solution type of the lowest energy soliton can change. Our numerical results are of relevance for the quantization of the classical soliton solutions.

\end{abstract}

\maketitle
\section{Introduction}

Hopf soliton solutions arise as topological solitons in the Skyrme-Faddeev model \cite{Faddeev:1975,Faddeev:1976pg} -- a nonlinear $O(3)$ sigma model in $(3+1)$-dimensional space-time whose Lagrangian is modified by an additional term quartic in its field derivatives. Extensive numerical simulations \cite{Faddeev:1996zj,Gladikowski:1996mb,Battye:1998zn,Battye:1998pe,Hietarinta:1998kt,Hietarinta:2000ci,Sutcliffe:2007ui} of the highly nonlinear classical field equations have revealed a very rich spectrum of solutions that are classified by their integer-valued Hopf charge. For Hopf charges up to 16 a variety of static, stable minimum-energy solutions with the structure of closed strings, twisted tori, linked loops, and knots have been identified. These stringlike solitons might be candidates to model glueball configurations \cite{Faddeev:1998eq} in QCD or may arise in two-component Bose condensates \cite{Babaev:2001zy,Jaykka:2006gf}. 

In this paper we investigate  the effect of isospin on classical Hopf soliton solutions. In analogy to the conventional $S\! U(2)$ Skyrme model we use the collective coordinate method \cite{Adkins:1983ya} to construct Hopf solitons of well-defined, nonzero isospin: We parametrize the isorotational zero modes of a Hopf configuration by collective coordinates, which are then taken to be time dependent. This gives rise to additional dynamical terms in the Hamiltonian, which can then be quantized following semiclassical quantization rules. A simplification that is often made in the literature \cite{Adkins:1983ya,Braaten:1988cc,Manko:2007pr,Krusch:2005bn} is to apply a simple adiabatic approximation to the (iso)rotational zero modes of the soliton by assuming that the soliton's shape is rotational frequency independent. The limitations of this rigid body approach were pointed out by several authors \cite{Braaten:1984qe,Battye:2005nx,Houghton:2005iu,Acus:2012st}. In this paper we perform numerical computations of isospinning Hopf solitons with Hopf charges up to 8 in the full three-dimensional classical field theory without applying the rigid body approximation and without imposing symmetry constraints on the isospinning Hopf configurations. It turns out that the Skyrme-Faddeev model with its rich topology of minimum-energy solutions, often of comparable energy, allows for ``transmutations'' when isospin is added and even for the formation of new, metastable Hopf solutions. 

This paper is organized as follows. In Sec.~\ref{Sec_SF} we briefly review the Skyrme-Faddeev model and describe how Hopf solitons acquire isospin within the collective coordinate approach. Then, in Sec.~ \ref{Sec_IC} we set up appropriate initial conditions which are used in Sec.~\ref{Sec_Stat} to compute Hopf configurations of zero isospin. The effect of isospin on these Hopf soliton solutions is studied in Sec.~\ref{Sec_Res}. We conclude with Sec.~\ref{Sec_Con}.

\section{Classically Isospinning Hopf Solitons}\label{Sec_SF}

The Lagrangian density of the Skyrme-Faddeev model \cite{Faddeev:1975} in $(3+1)$ dimensions takes in terms of the real three-component unit vector $\boldsymbol{\phi}=(\phi_1,\phi_2,\phi_3)$ the form
\begin{align}
\mathcal{L}&=\frac{1}{32\pi^2\sqrt{2}}\Bigg\{\partial_\mu\boldsymbol\phi\cdot\partial^\mu\boldsymbol\phi-\frac{1}{2}\left(\partial_\mu\boldsymbol\phi\times\partial_\nu\boldsymbol\phi\right)^2-V\left(\boldsymbol{\phi}\right)\Bigg\}\,.
\label{Lag_Hopf_phi}
\end{align}

To stabilize isospinning Hopf configurations we modified in (\ref{Lag_Hopf_phi}) the usual Skyrme-Faddeev model by adding a mass term $V\left(\boldsymbol{\phi}\right)$ to the $O(3)$ sigma model and Skyrme term. Here we will consider the following $S\! O(3)$ symmetry breaking potentials:
\begin{align}
V\left(\boldsymbol{\phi}\right)=\begin{cases}V_I\left(\boldsymbol{\phi}\right)&=2\mu^2\left(1-\phi_3\right)\,,\\
V_{II}\left(\boldsymbol{\phi}\right)&=\mu^2\left(1-\phi_3^2\right)\,,\end{cases}
\label{Lag_pot}
\end{align}
where $\mu$ is a rescaled mass parameter. The potential $V_I$ has one vacuum for $\phi_3=+1$, whereas $V_{II}$ has two vacua: for $\phi_3=+1$ and $\phi_3=-1$. The planar version of (\ref{Lag_Hopf_phi}) with $V=V_I$ corresponds to the old Baby Skyrme model \cite{Piette:1994ug}, and the one with $V=V_{II}$ reproduces the new Baby Skyrme model \cite{Kudryavtsev:1997nw,Weidig:1998ii}. The normalization in (\ref{Lag_pot}) is choosen so that for $\phi_3\rightarrow +1$ both potentials show the same asymptotic behaviour, explicitely given by $\mu^2(\phi_1^2+\phi_2^2)$.    

The Lagrangian (\ref{Lag_Hopf_phi}) admits topologically nontrivial, stringlike, finite-energy configurations due to the third homotopy group of the 2-sphere being nontrivial, $\pi_3(S^2)=\mathbb{Z}$. This can be seen as follows. A static finite-energy configuration requires the boundary condition $\boldsymbol{\phi}(t,\boldsymbol{x})\rightarrow(0,0,1)$ as $|\boldsymbol{x}|\rightarrow\infty$ for all time $t$.  Hence this boundary condition on the field $\boldsymbol{\phi}$ defines a mapping $\boldsymbol{\phi}: S^3\mapsto S^2$, and the field configurations can be classified topologically by the homotopy group $\pi_3(S^2)=\mathbb{Z}$. The topological invariant associated with each static field configuration is known as the Hopf charge $N$. It can be interpreted geometrically as the linking number of two loops obtained as the preimages of any two generic distinct points on the target 2-sphere. The position curve of the soliton is defined as the set of points where the field is as far as possible from the boundary vacuum value $\boldsymbol{\phi_\infty}=(0,0,1)$. Thus it is given by the preimage of the point $-\boldsymbol{\phi_\infty}$, which is antipodal to the vacuum value. When we visualize Hopf solitons' position curves, we usually display for clarity tubelike isosurfaces with $(0,0,-1+\delta)$, where $\delta$ is chosen to be small. Similarly the linking curve can be illustrated graphically by plotting an isosurface of the preimage of the vector $(-1+\delta,0,0)$.   

The overall factor $1/32\pi^2/\sqrt{2}$ in (\ref{Lag_Hopf_phi}) is motivated by Ward's conjecture \cite{Ward:1998pj} that with the normalization (\ref{Lag_Hopf_phi}) the Vakulenko-Kapitanski lower bound \cite{Kundu:1982bc,Vakulenko:1979} on the energy $M_N$  of a Hopf configuration with charge $N$ is given by
\begin{align}
M_N\geq c N^{3/4}\,,\quad\text{where }\,c=1\,.
\label{Ward_bound}
\end{align}
The topological bound (\ref{Ward_bound}) has been shown to be compatible with fully three-dimensional numerical simulations carried out in the massless Skyrme-Faddeev model \cite{Battye:1998zn,Battye:1998pe,Hietarinta:2000ci} and in the massive one \cite{Foster:2010zb} with potential $V_{I}$ included.

The Skyrme-Faddeev model (\ref{Lag_Hopf_phi}) can be expressed in analogy to the conventional $S\! U(2)$ Skyrme model \cite{Skyrme:1961vq} in terms of the $S\! U(2)$-valued Hermitian scalar field $U(t,\boldsymbol{x})=\boldsymbol{\phi}\cdot\boldsymbol{\tau}$. 
The ansatz for the dynamical soliton field adopted in the collective coordinate quantization  \cite{Adkins:1983ya,Braaten:1988cc,Manko:2007pr} is given by
\begin{align}
\widehat{U}(\boldsymbol{x},t)=A_1(t)U_0(\boldsymbol{x})A_1^\dagger(t)\,,
\label{U_dyn}
\end{align}
where we have promoted the collective coordinate $A_1\in S\! U(2)$ to a time-dependent dynamical variable and ignored the translational and rotational degrees of freedom. $A_1(t)$ describes the isorotational fluctuations about the classical minimum-energy solution $U_0(\boldsymbol{x})$. Substituting (\ref{U_dyn}) in (\ref{Lag_Hopf_phi}) and defining the body-fixed angular velocities via $a_j=-\i\text{Tr}(\tau_jA_1^\dagger\dot{A}_1)$ the Skyrme-Faddeev Lagrangian takes the form 
\begin{align}
L=\frac{1}{2}a_iU_{ij}a_j-M_N\,,
\end{align}
where the Hopf soliton mass $M_N$ is given by 
\begin{align}
M_N&=\frac{1}{32\pi^2\sqrt{2}}\int\Bigg\{\partial_i\boldsymbol\phi\cdot\partial_i\boldsymbol\phi+\frac{1}{2}\left[\left(\partial_i\boldsymbol\phi\cdot\partial_i\boldsymbol\phi\right)^2-\left(\partial_i\boldsymbol\phi\cdot\partial_j\boldsymbol\phi\right)\left(\partial_i\boldsymbol\phi\cdot\partial_j\boldsymbol\phi\right)\right]+V\left(\boldsymbol{\phi}\right)\Bigg\}\,\text{d}^3x\,,
\label{Hopf_mass_phi}
\end{align}
and the moment of inertia tensors is
\begin{subequations}
\begin{align}
U_{ij}&=\frac{1}{16\pi^2\sqrt{2}}\int\, \left(\boldsymbol{\phi}^2\delta_{ij}-\phi_i\phi_j\right)\left(1+\partial_k\boldsymbol{\phi}\cdot\partial_k\boldsymbol{\phi}\right)-\left(\boldsymbol{\phi}\times\partial_k\boldsymbol{\phi}\right)_i\left(\boldsymbol{\phi}\times\partial_k\boldsymbol{\phi}\right)_j\,\text{d}^3x\,.\label{Hopf_Inertia_phi_u}
\end{align}
\label{Hopf_Inertia_phi}
\end{subequations}
The momentum conjugate to $a_i$ is the body-fixed isorotation angular momentum $K_i$ defined via
\begin{align}
K_i&=\frac{\partial L}{\partial a_i}=U_{ij}a_j\,.
\end{align}

In this article, we choose the $z$ axis as our rotation axis. Using gradient-based mehods we search for Hopf configurations $\boldsymbol{\phi}$ of a given topological charge $N$, which minimize
\begin{align}
-L=M_N-\frac{1}{2}U_{33}\omega^2\,,
\label{E_min}
\end{align}
where  the rotation frequency $\omega= a_3$ is calculated at each time step for a fixed $|K|$ as follows 
\begin{align}
\omega=\frac{|K|}{U_{33}}\,.
\end{align}

\section{Initial Conditions}\label{Sec_IC}

We create suitable initial field configurations with nontrivial Hopf charge $N$ by using the approach presented in Ref.~\cite{Sutcliffe:2007ui}. The basic idea is to approximate the Hopf configuration by rational maps $W:\,S^3\mapsto \mathbb{C}\mathbb{P}^1$, that is, a mapping from the three-sphere to the complex projective line. This approach enables us to set up initial conditions for knotted, linked and axial Hopf configurations with energies reasonably close to the suspected minimum energy solutions. These initial conditions can then be relaxed using a modified version of the energy minimization algorithm \cite{Battye:2001qn} originally designed to study Skyrmion solutions.

First we compactify $\mathbb{R}^3$ to a unit 3-sphere $S^3\in \mathbb{C}^2$ via a degree one spherically equivariant map given by
\begin{align}
\left(Z_1,Z_0\right)=\left(\frac{x_1+\i x_2}{r}\sin f,\cos f + \i \frac{\sin f}{r}x_3\right)\,,
\label{Map_Z}
\end{align}
where $(x_1,x_2,x_3)\in \mathbb{R}^3$, $r^2=x_1^2+x_2^2+x_3^2$ and $(Z_1,\,Z_0)$ are complex coordinates on the unit 3-sphere (with $|Z_1|^2+|Z_0|^2=1$). Here $f(r)$ is a monotonically decreasing profile function with boundary conditions $f(0)=\pi$ and $f(\infty)=0$. In our simulations we use a simple linear profile function $f(r)=\pi(r_{\text{max}}-r)/L$ for $r<r_{\text{max}}$ and $f(r)=0$ for $r\geq{r_{\text{max}}}$ with $r_{\text{max}}=6$. Approximate Hopf solutions can be obtained by writing the stereographic projection of the field $\boldsymbol{\phi}$
\begin{align}
W=\frac{\phi_1+\i\phi_2}{1+\phi_3}\,,
\end{align}
as a rational function of the complex variables $Z_1$ and $Z_0$
\begin{align}
W=\frac{p(Z_1,\,Z_0)}{q(Z_1,\,Z_0)}\,,
\end{align}
where $p$ and $q$ are polynomials in $Z_1$ and $Z_0$.

There are three different solution types which will be used as initial field configurations for our energy relaxation simulations:
\begin{itemize}
\item Toroidal fields of solution type $\mathcal{A}_{n,m}$ can be obtained by setting
\begin{align}
W&=\frac{Z_1^n}{Z_0^m}\,,
\label{Axial_rational_Hopf}
\end{align}
where $n,\,m\in \mathbb{Z}$. The integer pair  $(n,m)$ counts the angular windings around the two cycles of the torus.
An axially symmetric Hopf configuration of the type $\mathcal{A}_{n,m}$ can be described \cite{Battye:1998zn,Sutcliffe:2007ui} by a baby Skyrmion solution with winding number $m$, which is embedded in the $(3+1)$-dimensional Skyrme-Faddeev model (\ref{Lag_Hopf_phi}) along a closed curve and with its internal phase rotated through an angle $2\pi n$ as it travels around the circle once. The Hopf charge $N$ associated with such an unlinked Hopf configuration (\ref{Axial_rational_Hopf}) is given by $N=nm$. 

\item $(a,b)$-torus knots $\mathcal{K}_{ab}$ are described by the mapping
\begin{align}
W=\frac{Z_1^\alpha Z_0^\beta}{Z_1^a+Z_0^b}\,,
\label{Knot_rational_Hopf}
\end{align}
where $\alpha$ is a positive integer, $\beta$ is a non-negative integer and $a,b$ are coprime positive integers with $a>b$. The rational map (\ref{Knot_rational_Hopf}) generates a knot lying on the surface of an unknotted torus and winding $a$ and $b$ times about the torus circumferences. Fields of type $\mathcal{K}_{a,b}$ have topological charge  $N=\alpha b+\beta a$ \cite{Sutcliffe:2007ui}. 

\item Linked Hopf initial configurations of the type $\mathcal{L}^{\alpha,\beta}_{p,q}$  can be constructed, when the denominator of (\ref{Knot_rational_Hopf}) is reducible. Following the notation of \cite{Sutcliffe:2007ui}, $p$ and $q$ label the charges of the two disconnected components that form the link, and the additional linking number of each component due to its linking with the other is denoted by the superscripts $\alpha$ and $\beta$. The total Hopf charge of a field $\mathcal{L}^{\alpha,\beta}_{p,q}$ is  $N=p+q+\alpha+\beta$. In particular, in this paper we will use the rational map 
\begin{align}
W&=\frac{Z_1^{n+1}}{Z_1^2-Z_0^2}=\frac{Z_1^n}{2\left(Z_1-Z_0\right)}+\frac{Z_1^n}{2\left(Z_1+Z_0\right)}\,,
\label{Link_rational_Hopf}
\end{align}
to produce smooth initial linked configurations of solution type $\mathcal{L}^{1,1}_{n,n}$ and Hopf charge $N=2n+2$. 
\end{itemize}

In the following section, we compute minimum-energy Hopf solutions for potential $V_I$ and $V_{II}$ using a relaxation algorithm with initial conditions constructed from the rational maps (\ref{Axial_rational_Hopf}), (\ref{Knot_rational_Hopf}) and from linked configurations like e.g., (\ref{Link_rational_Hopf}). To avoid saddle point solutions of the Skyrme-Faddeev energy functional $M_N$, we explicitely add, in a similar way to  \cite{Battye:1998zn}, symmetry-breaking, nonaxial perturbations to our initial conditions.

\section{Relaxed Hopf Soliton Solutions}\label{Sec_Stat}
To find the stationary points of the energy functional $M_N$, we solve the associated Euler-Lagrange equations numerically. The field equations can be implemented analogous to Ref.~\cite{Battye:2001qn}
\begin{align}
M\ddot{\boldsymbol{\phi}}-\boldsymbol{\alpha}\left(\dot{\boldsymbol{\phi}},\partial_i\boldsymbol{\phi},\partial_i\dot{\boldsymbol{\phi}},\partial_i\partial_j\boldsymbol{\phi}\right)-\lambda\boldsymbol{\phi}+\epsilon\dot{\boldsymbol{\phi}}=0\,,
\label{Num_Flow}
\end{align}
where $M$ is a symmetric matrix. The dissipation $\epsilon$ in (\ref{Num_Flow}) is added to speed up the relaxation process, and the Lagrange multiplier $\lambda$ imposes the unit vector constraint $\boldsymbol{\phi}\cdot\boldsymbol{\phi}=1$. We do not present the full field equations here since they are cumbersome and not particularly enlightening. The initial configuration is then evolved according to the flow equations (\ref{Num_Flow}). Kinetic energy is removed periodically by setting $\dot{\boldsymbol{\phi}}=0$ at all grid points. All the simulations presented in the following use fourth order spatial differences on grids with $(201)^3$ points, a spatial grid spacing $\Delta x=0.1$, and time step size $\Delta t=0.01$. The dissipation is set to $\epsilon=0.5$, and we choose the rescaled mass parameter $\mu=1$ throughout this paper.

\begin{figure}[h]
\centering
\subfigure[$1\mathcal{A}_{1,1}$]{\includegraphics[totalheight=1.5cm]{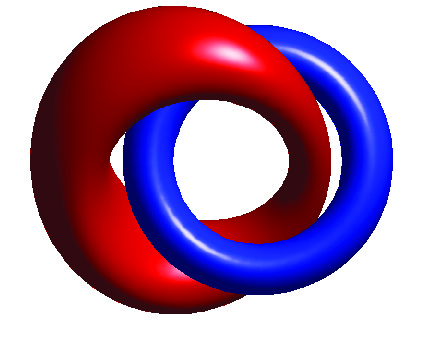}}
\subfigure[$2\mathcal{A}_{2,1}$]{\includegraphics[totalheight=1.5cm]{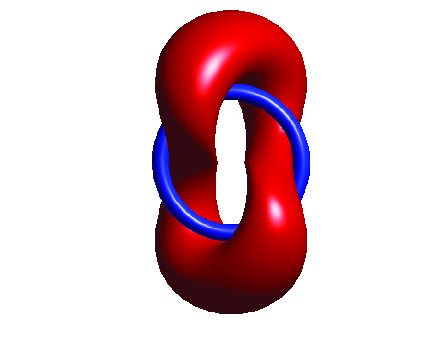}}
\subfigure[$3\mathcal{\widetilde{A}}_{3,1}$]{\includegraphics[totalheight=1.5cm]{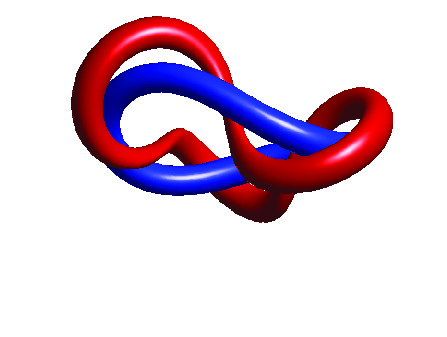}}
\subfigure[$3\mathcal{A}_{3,1}$]{\includegraphics[totalheight=1.5cm]{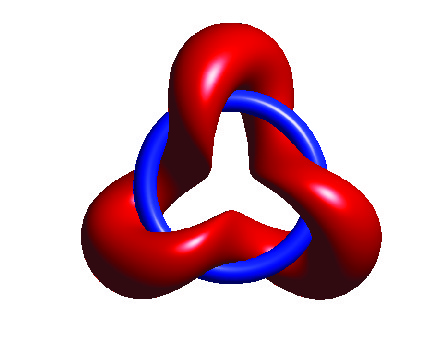}}
\subfigure[$4\mathcal{\widetilde{A}}_{4,1}$]{\includegraphics[totalheight=1.5cm]{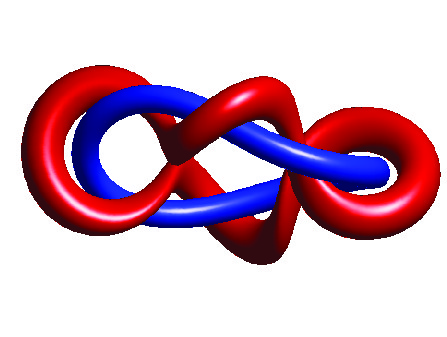}}
\subfigure[$4\mathcal{A}_{2,2}$]{\includegraphics[totalheight=1.5cm]{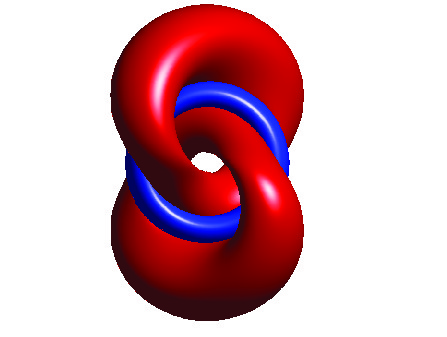}}
\subfigure[$4\mathcal{A}_{4,1}$]{\includegraphics[totalheight=1.5cm]{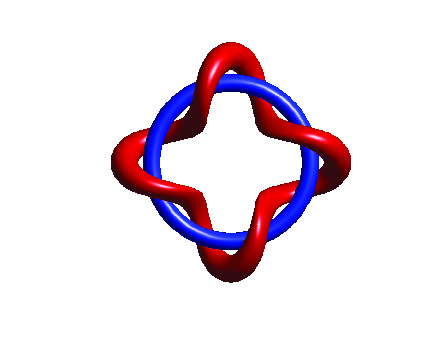}}\\
\subfigure[$5\mathcal{L}^{1,1}_{1,2}$]{\includegraphics[totalheight=1.5cm]{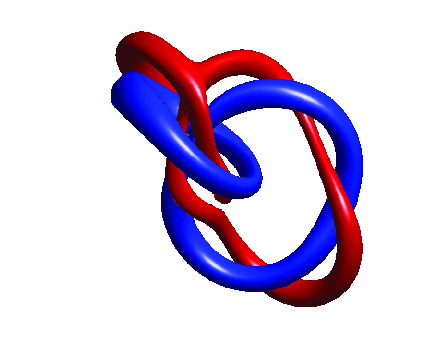}}
\subfigure[$5\widetilde{\mathcal{A}}_{5,1}$]{\includegraphics[totalheight=1.5cm]{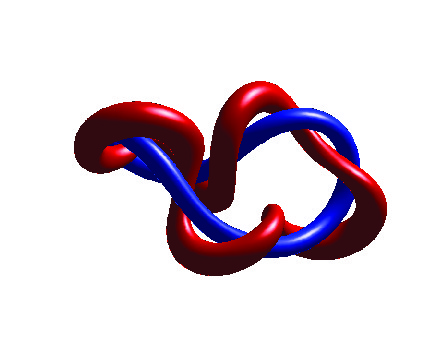}}
\subfigure[$5\mathcal{A}_{5,1}$]{\includegraphics[totalheight=1.5cm]{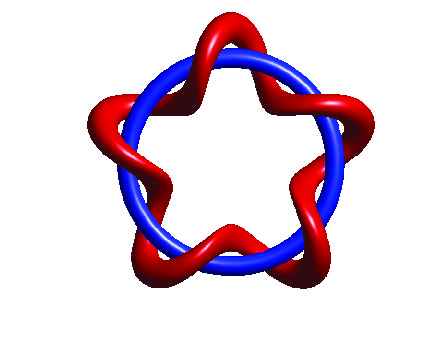}}
\subfigure[$6\mathcal{A}_{3,2}$]{\includegraphics[totalheight=1.5cm]{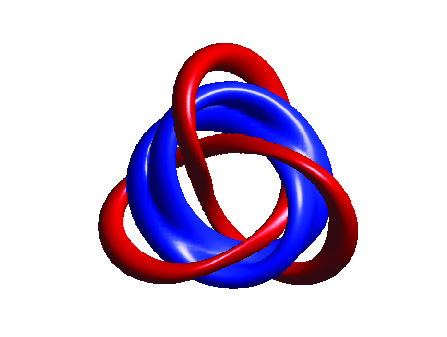}}
\subfigure[$6\mathcal{L}^{1,1}_{2,2}$]{\includegraphics[totalheight=1.5cm]{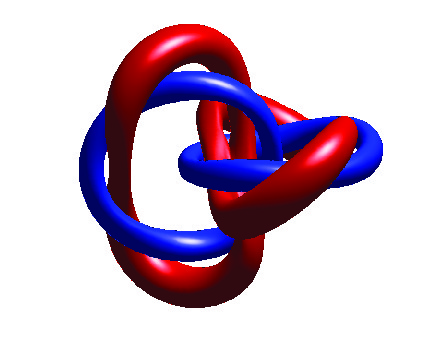}}
\subfigure[$6\mathcal{L}^{1,1}_{3,1}$]{\includegraphics[totalheight=1.5cm]{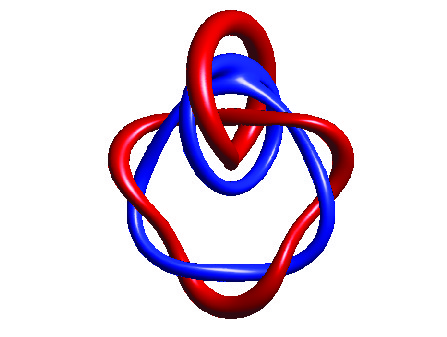}}
\subfigure[$6\mathcal{\widetilde{A}}_{6,1}$]{\includegraphics[totalheight=1.5cm]{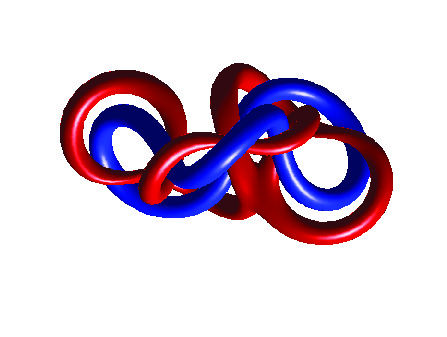}}\\
\subfigure[$6\mathcal{A}_{6,1}$]{\includegraphics[totalheight=1.5cm]{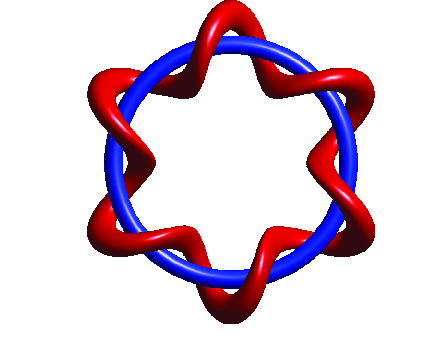}}
\subfigure[$7\mathcal{K}_{3,2}$]{\includegraphics[totalheight=1.5cm]{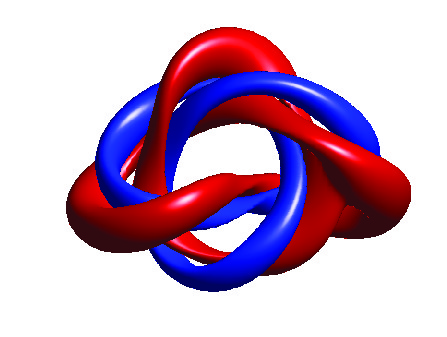}}
\subfigure[$7\mathcal{K}_{2,3}$]{\includegraphics[totalheight=1.5cm]{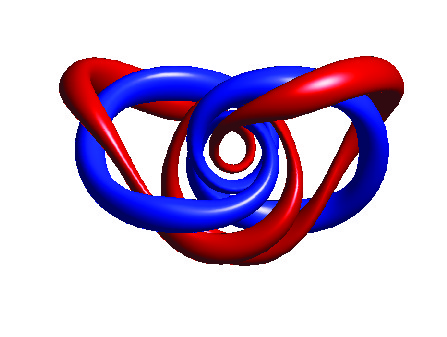}}
\subfigure[$7\mathcal{\widetilde{A}}_{7,1}$]{\includegraphics[totalheight=1.5cm]{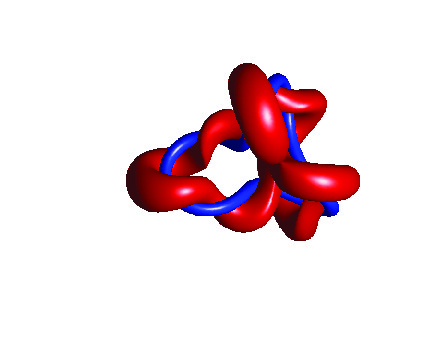}}
\subfigure[$8\mathcal{\widetilde{A}}_{4,2}$]{\includegraphics[totalheight=1.5cm]{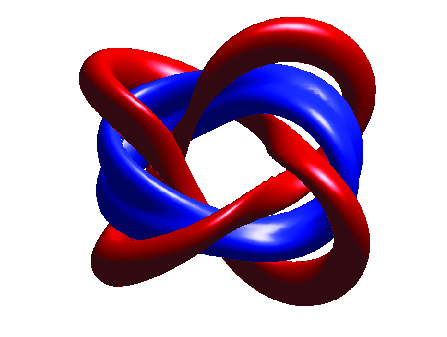}}
\subfigure[$8\mathcal{K}_{3,2}$]{\includegraphics[totalheight=1.5cm]{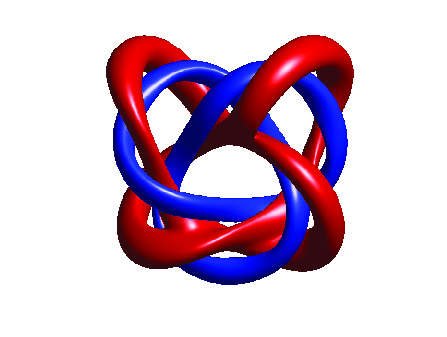}}
\subfigure[$8\mathcal{\widetilde{A}}_{8,1}$]{\includegraphics[totalheight=1.5cm]{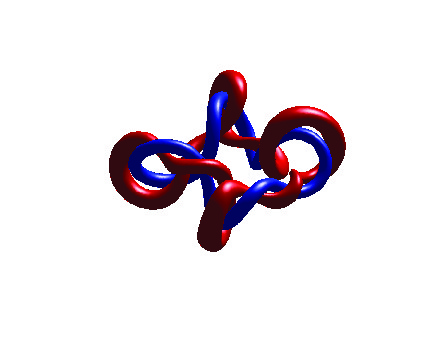}}
\caption{Position (blue tube) and linking (red tube) curve for $\mu=1$ Hopf solitons with Hopf charge $1\leq N\leq 8$ and potential $V_I$. We label each configuration by its Hopf charge $N$ and its solution type. The corresponding energy values can be found in Table \ref{Tab_Hopf_en}. }
\label{Fig_Hopf_link}
\end{figure}

A summary of our relaxed configurations is given in Table~\ref{Tab_Hopf_en}. Each initial configuration is listed together with the final Hopf configuration it evolves to.      
In Fig.~\ref{Fig_Hopf_link} we display the linking structure of the minimum-energy configurations of charge $1\leq N \leq8$ obtained for potential $V_I$. Here, we visualize the field configurations by plotting isosurfaces of the points $(0,0,-\epsilon)$ and $(-\epsilon,0,0)$ with $\epsilon=0.8$. Our calculations with potential $V_{II}$ produce the same Hopf solution types as for potential $V_I$, the main difference being that the solitons are more compact. The minimal energy solutions  of both massive models are very similar to the massless ones \cite{Battye:1998zn,Battye:1998pe,Hietarinta:1998kt,Hietarinta:2000ci,Sutcliffe:2007ui}.

\begin{table}[!h]
\caption{All initial conditions and final $\mu=1$ Hopf configurations with their respective energies. $M_I$ and $M_{II}$ denote the soliton energy with potential $V_I$ and $V_{II}$ included, respectively. The superscript ``$\text{pert.}$'' indicates that we applied nonaxial perturbations to the initial configuration. $(\star)$ These configurations correspond to global energy minima for given Hopf charge $N$. Recall that energies are given in units of $1/32\pi^2/\sqrt{2}$. }
\begin{ruledtabular}
\begin{tabular}{cccccccc}
$N$  & initial   &final & $M_I$ &  $M_I/N^{{3}/{4}}$ &$M_{II}$ &  $M_{II}/N^{{3}/{4}}$    \\\hline
1    & $\mathcal{A}_{1,1}$   &$\mathcal{A}_{1,1}$& $1.438^{\star}$   & 1.438& $1.373^{\star}$ &1.373\\
2    & $\mathcal{A}_{2,1}$   &$\mathcal{A}_{2,1}$& $2.287^{\star}$   & 1.359&  $2.188^{\star}$ &1.300\\
3    & $\mathcal{A}^{\text{pert.}}_{3,1},\mathcal{K}^{\text{pert.}}_{2,1}$   &  $\mathcal{\widetilde{A}}_{3,1}$ & $3.173^\star$     & 1.391 & $3.041^{\star}$&1.334   \\
     & $\mathcal{A}_{3,1}$    & $\mathcal{{A}}_{3,1}$ & 3.178  & 1.394    & 3.048 &1.337 \\
4    & $\mathcal{A}^{\text{pert.}}_{4,1}, \mathcal{K}^{\text{pert.}}_{2,1}, \mathcal{K}^{\text{pert.}}_{4,1}$    &  $\mathcal{\widetilde{A}}_{4,1}$ &   $4.034^{\star}$   & 1.426& 3.862 &1.365  \\
    &  $\mathcal{A}_{2,2}, \mathcal{L}^{1,1}_{1,1}$   & $ \mathcal{{A}}_{2,2}$ & $4.060$    & 1.435   & $3.844^{\star}$ &1.359\\
    &  $\mathcal{A}_{4,1}$   & $\mathcal{{A}}_{4,1}$ & 4.104   & 1.450& 3.943  &1.394 \\
5    & $\mathcal{K}^{\text{pert.}}_{3,2}$   &  $\mathcal{L}^{1,1}_{1,2}$ &  $4.871^{\star}$  & 1.456 & $4.549^{\star}$ &1.360\\
    &  $\mathcal{K}^{\text{pert.}}_{4,1}$   & $\mathcal{\widetilde{A}}_{5,1}$ &  4.890 & 1.462 & 4.685& 1.401\\
    &  $\mathcal{A}_{5,1}$   & $\mathcal{A}_{5,1}$ &  5.047  & 1.509 & 4.756& 1.422\\
6    & $\mathcal{K}^{\text{pert.}}_{3,2}, \mathcal{K}^{\text{pert.}}_{4,2},\mathcal{A}_{3,2}$   &  $\mathcal{A}_{3,2}$ &  $5.402^{\star}$  &1.409 & $5.134^{\star}$ &1.339\\
    &  $\mathcal{K}^{\text{pert.}}_{2,2}$   & $\mathcal{L}^{1,1}_{2,2}$ &  5.455  & 1.422 & 5.198 &1.355\\
    &  $\mathcal{L}^{1,1}_{3,1}$   & $\mathcal{L}^{1,1}_{3,1}$ & 5.556   & 1.449 & 5.285  &1.378\\
    &  $\mathcal{K}^{\text{pert.}}_{5,1}$   & $\mathcal{\widetilde{A}}_{6,1}$ &  5.642  & 1.471 &5.481&1.429 \\
    &  $\mathcal{A}_{6,1}$   & $\mathcal{A}_{6,1}$ &  6.001  & 1.565& 5.541 &1.445\\
7    &      $\mathcal{K}^{\text{pert.}}_{4,3}, \mathcal{K}^{\text{pert.}}_{5,2}$ & $\mathcal{K}_{3,2}$ &  $6.138^{\star}$   &1.426 & $5.822^{\star}$&1.352\\
    &      $\mathcal{K}_{2,3}$ &$\mathcal{K}_{2,3}$ & 6.450   & 1.498& 6.129&1.424\\
    &      $\mathcal{A}^{\text{pert.}}_{7,1}$ &$\mathcal{\widetilde{A}}_{7,1}$ & 6.587 &1.530 & 6.294&1.462\\
8   &     $\mathcal{A}_{4,2},\mathcal{L}^{2,2}_{2,2},\mathcal{L}^{1,1}_{3,3},\mathcal{K}^{\text{pert.}}_{3,4}$ &$\mathcal{\widetilde{A}}_{4,2}$ &  $6.747^{\star}$  &1.418 & $6.414^{\star}$&1.348\\
    &      $\mathcal{K}_{3,2}, \mathcal{K}^{\text{pert.}}_{5,2}$  &$\mathcal{K}_{3,2}$&  6.754  & 1.419 & 6.433 &1.352\\
    &      $\mathcal{A}^{\text{pert.}}_{8,1}$ &$\mathcal{\widetilde{A}}_{8,1}$ &  7.201  &1.513 & 6.844 &1.438\\
\end{tabular}
\end{ruledtabular}
\label{Tab_Hopf_en}
\end{table}

Relaxing (\ref{Axial_rational_Hopf}) with $n=m=1$ reproduces the $\mathcal{A}_{1,1}$ static Hopf configuration, which has for $V_I$ an energy $M_1=1.438$ and a moment of inertia $U_{33}$=0.500, and this agrees well with $M_1=1.421$ stated in Ref.~\cite{Foster:2010zb}. For comparison, substituting a spherically symmetric hedgehog form $U(\boldsymbol{x})=\exp(\i f(r)\,\boldsymbol{\widehat{r}}\cdot\boldsymbol{\tau})$ in (\ref{Hopf_mass_phi}) and minimzing the energy with respect to the profile function $f$ gives for the 1-Hopf soliton solution an energy $M_1=1.452$ and a moment of inertia $U_{33}=0.502$. The minimal energy $N=2$ Hopf solitons are of the type $\mathcal{A}_{2,1}$ -- axially symmetric configurations with the linking curve twisted two times around the position curve. Applying nonaxial perturbations to an $\mathcal{A}_{3,1}$ or  a $\mathcal{K}_{2,1}$ initial configuration, we find the $\mathcal{\widetilde{A}}_{3,1}$ 3-Hopf soliton solution to be of lowest energy for both potential choices. Here, the tilde indicates that the position curve is not lying completely  in the plane but it is bent. For completeness, we also include in Tab. \ref{Tab_Hopf_en} and in Fig. \ref{Fig_Hopf_link} minimal-energy configurations of solution type $N\mathcal{A}_{N,1}$. These axial solutions are known to be unstable for $N\geq3$ \cite{Battye:1998zn,Battye:1998pe}. Taking perturbed axially symmetric $\mathcal{A}_{4,1}$ and knotted $\mathcal{K}_{2,1},\mathcal{K}_{4,1}$ configurations as our initial conditions, we identify the bent axial solution $\mathcal{\widetilde{A}}_{4,1}$ as the global energy minimimum for $N=4$ and potential $V_I$ \cite{Foster:2010zb}. The charge-4 configuration $\mathcal{A}_{2,2}$ (created from axial and linked initial conditions) and $\mathcal{A}_{4,1}$ are local energy minima. However, for potential $V_{II}$ the minima swap with $\mathcal{A}_{2,2}$ becomes the lowest minimal-energy charge-4 soliton solution. For $N=5$ the minimal configuration in both massive models is a link of type $\mathcal{L}^{1,1}_{1,2}$, which we obtained by relaxing a perturbed trefoil knot $\mathcal{K}_{3,2}$. The charge-5 bent solution $\mathcal{\widetilde{A}}_{5,1}$ and the toroidal $\mathcal{A}_{5,1}$ seem to be metastable local minima. For $N=6$ we find using a variety of initial conditions that the $\mathcal{A}_{3,2}$ configuration has minimal energy, whereas the links $\mathcal{L}^{1,1}_{2,2},\,\mathcal{L}^{1,1}_{3,1}$, the bent unknot $\mathcal{\widetilde{A}}_{6,1}$, and the rotationally symmetric unknot $\mathcal{A}_{6,1}$ are only local minima \cite{Foster:2010zb}. This differs from the massless Skyrme-Faddeev model where the link $\mathcal{L}^{1,1}_{2,2}$ is the minimal-energy charge-6 soliton. Similar to the massless case, the trefoil knot $\mathcal{K}_{3,2}$ turns out to be the global minimum for $N=7$ in the massive models. Charge-7 Hopf solutions like the $\mathcal{K}_{2,3}$ knot  and the bent unknot $\mathcal{\widetilde{A}}_{7,1}$ represent local minima. Finally, for $N=8$ we identify $\mathcal{\widetilde{A}}_{4,2}$ as the minimal-energy solution. For potential $V_I$ the trefoil knot $8\mathcal{K}_{3,2}$ can be seen within the numerical accuracy as an almost energy-degenerate state. The link $\mathcal{L}^{1,1}_{3,3}$ which is the minimal-energy solution type in the massless model relaxes to $\mathcal{\widetilde{A}}_{4,2}$.
   
\begin{figure}[htb]
\centering
\subfigure[Normalized energies for Hopf solitons in the Faddeev-Skyrme model modified by  potential  $V_{I}$ vs Hopf charge $N$.]{\includegraphics[totalheight=6cm]{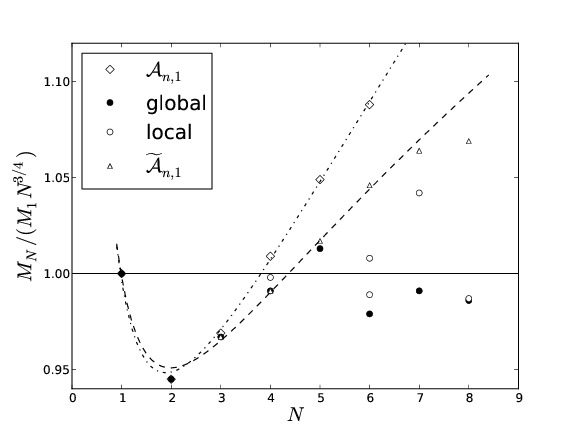}}
\subfigure[Same as (a) but for potential $V_{II}$.]{\includegraphics[totalheight=6cm]{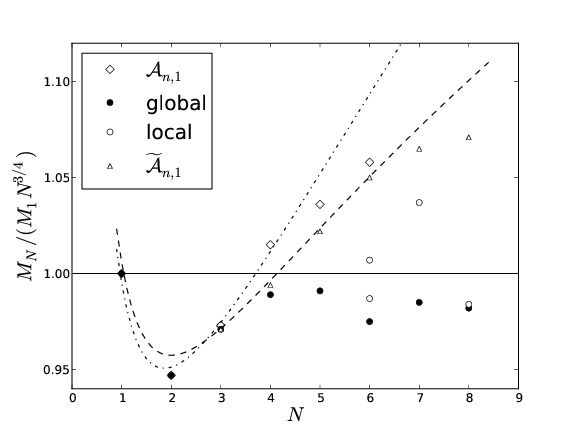}}
\caption{The normalized energies $M^{\star}_N=M_N/\left(M_1N^{3/4}\right)$ for different minimal energy, massive Hopf soliton solutions as a function of the Hopf charge $N$. The mass parameter $\mu$ is chosen to be 1. Here, our global minima for $1\leq N \leq 8$ are represented by filled circles ($\bullet$), bent unknots $N\mathcal{\widetilde{A}}_{N,1}$ by triangles ($\blacktriangle$), rotationally symmetric unknots $N\mathcal{A}_{N,1}$ by diamonds ($\diamond$), and the remaining local energy minima are displayed as open circles ($\circ$). The dashed line shows our linear fit to the bent unknots $N=1-5$ ($\mathcal{A}_{1,1},\,\mathcal{A}_{2,1}$ included), whereas the dash-dotted line represents a quadratic fit to the rotationally symmetric unknots $N=1-6$. The expected $N^{3/4}$ power growth is represented by the horizontal line. The corresponding plots for the massless Faddeev-Skyrme model can be found in Refs.~\cite{Hietarinta:2000ci} and \cite{Sutcliffe:2007ui}.}
\label{Fig_Ebound}
\end{figure}

In Fig.~\ref{Fig_Ebound} we show in analogy to Ref.~\cite{Hietarinta:2000ci} the normalized minimum energies $M^{\star}_N=M_N/\left(M_1N^{3/4}\right)$ for both potential choices. In both cases the energies of the ground-state Hopf configurations (filled circles) follow $M_N\propto N^{3/4}$. As already pointed out in Ref.~\cite{Hietarinta:2000ci} for the massless case, the energies for the $2\mathcal{A}_{2,1}$ configurations in the massive models are particularly low compared to the standard level. We verify in Fig.~\ref{Fig_Ebound} that the normalized energies of the bent configurations $N\mathcal{\widetilde{A}}_{N,1}$ with $N=3-5$ are well described by the linear \cite{Miettinen:1999ir} fits $M_N/M_1=0.39+0.6N$ and $M_N/M_1=0.40+0.6N$ for $V_I$ and $V_{II}$, respectively. A very similar fit ($M_N/M_1=0.36+0.65N$) is given in Ref.~\cite{Hietarinta:2000ci} for the bent unknots in the massless Skyrme-Faddeev model. For the planar configurations $N\mathcal{A}_{N,1}$ with $N=1-6$ we obtain $M_N/M_1=0.415+0.5719N+0.009N^2$ for $V_I$ and $M_N/M_1=0.4348+0.5501N+0.01475N^2$ for $V_{II}$ and $N=1-4$. The corresponding quadratic fit for massless rotationally symmetric unknots is given as $M_N/M_1=0.39+0.59N+0.015N^2$ in Ref.~\cite{Hietarinta:2000ci}.

\section{Numerical Results on Classically Isospinning Hopf Solitons}\label{Sec_Res}

In this section, we present the results of our energy minimization simulations of isospinning Hopf solitons with charges $N$ up to 8. The variational equations derived from (\ref{E_min}) are implemented in analogy to (\ref{Num_Flow}), where we include in $\boldsymbol{\alpha}$ the isorotational extra terms. We use the configurations obtained in the previous sections as our start configurations for vanishing angular momentum ($K=0$) and increase $K$ in a stepwise manner. All simulation parameters are chosen as stated in Sec.~\ref{Sec_Stat}. In particular, we use the mass parameter $\mu=1$ and work on grids containing $(201)^3$ lattice points with a lattice spacing $\Delta x=0.1$. If not stated otherwise, we use $V=V_I$ as our potential term in (\ref{Lag_Hopf_phi}).    

Note that for $\mu\leq 1$ there exists a maximal frequency $\omega_{max}=\mu$ beyond which no stable isospinning Hopf soliton solution exists. This upper limit follows from the stability analysis of the linearized Euler-Lagrange equations derived from (\ref{E_min}).  

\subsection{Low Charge Hopf Solitons: $1\leq N\leq 3$}

\begin{figure}[htb]
\centering
\subfigure[$E_\text{tot}$ as a function of $\omega$]{\includegraphics[totalheight=4.cm]{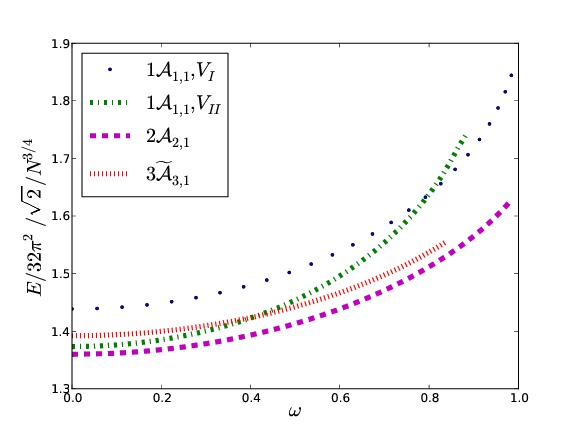}}
\subfigure[$E_\text{tot}$ as a function of $K$]{\includegraphics[totalheight=4.cm]{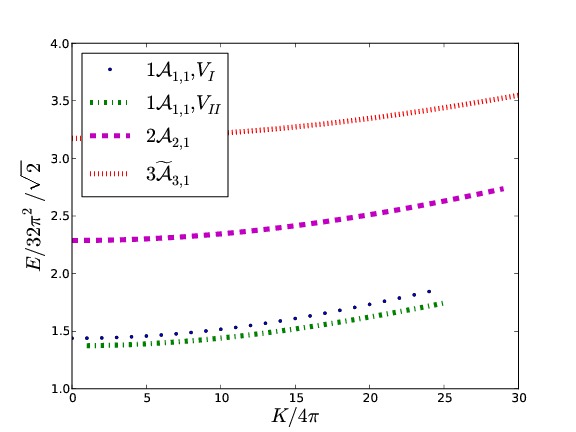}}
\subfigure[$U_{33}$ as a function of $\omega$]{\includegraphics[totalheight=4.cm]{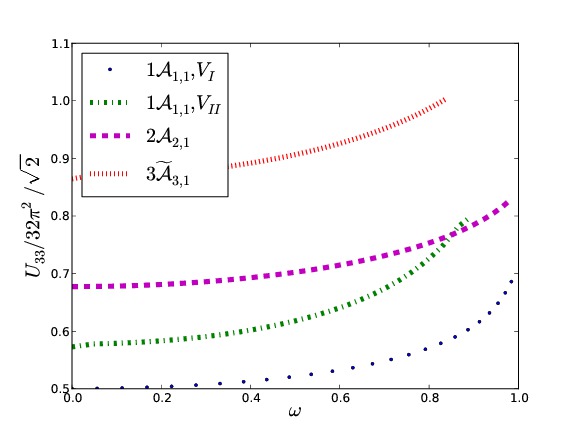}}
\caption{Energy and moment of inertia of isospinning Hopf solitons with $N=1-3$. For $N=1$ the results for both potential choices are shown. }
\label{Ew_isospin_N1N3}
\end{figure}

We show in Fig.~\ref{Ew_isospin_N1N3} the total energy $E_{\text{tot}}$ as a function of the rotation frequency $\omega$ and the angular momentum $K$ for isospinning Hopf solitons (of type $1\mathcal{A}_{1,1},\,2\mathcal{A}_{2,1},3\mathcal{\widetilde{A}}_{3,1}$) with charges up to 3. The corresponding plots for the moment of inertia $U_{33}$ as function of $\omega$ are also presented. For all these configurations the solution type of the isospinning soliton is the same as the one in the static case, only the soliton's size grows with $\omega$ and $K$. As expected, the energies and moment of inertia diverge for $\omega=\mu$.

\begin{figure}[htb]
\centering
\subfigure[$E_\text{tot}$ as a function of $\omega$]{\includegraphics[totalheight=4.cm]{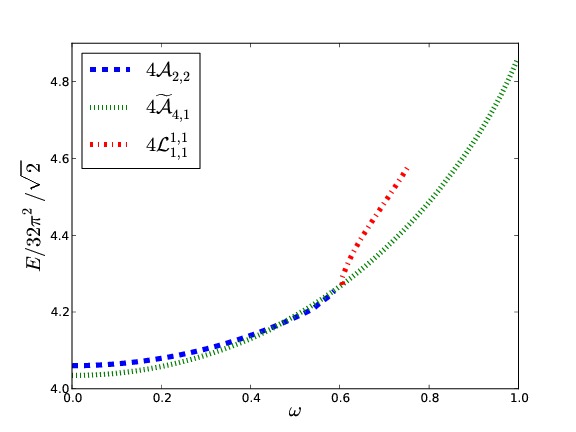}}
\subfigure[$E_\text{tot}$ as a function of $K$]{\includegraphics[totalheight=4.cm]{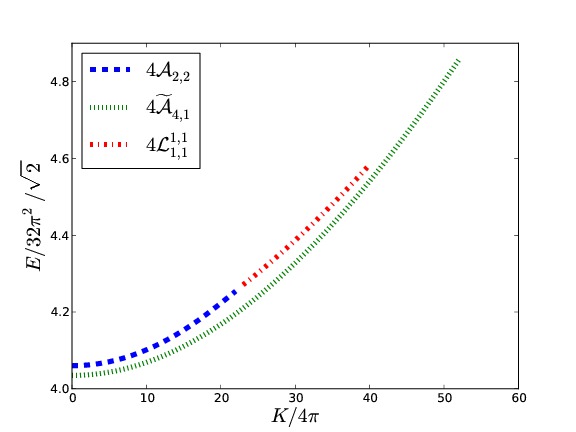}}
\subfigure[$U_{33}$ as a function of $\omega$]{\includegraphics[totalheight=4.cm]{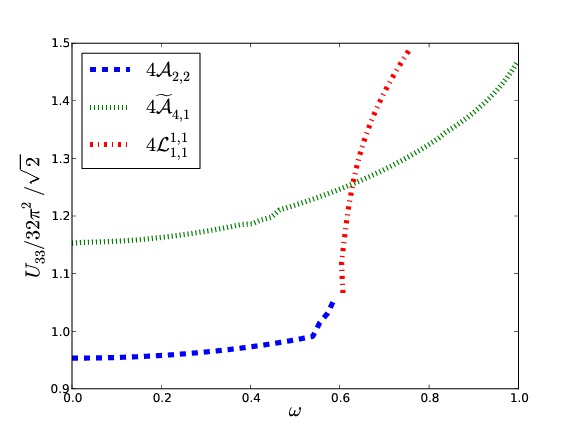}}
\caption{Total energy $E_\text{tot}$ and moment of inertia $U_{33}$ of isospinning 4-Hopf solitons calculated with potential $V_{I}$. The isospinning $4\mathcal{A}_{2,2}$ soliton (blue curve) deforms into $4\mathcal{L}^{1,1}_{1,1}$ (red curve). The transition occurs at $\omega\approx0.606,\,K\approx 23$. The bent configuration $4\mathcal{\widetilde{A}}_{4,1}$ (green curve) exists for all $\omega\in[0,1)$ and its size is growing with $\omega$.}
\label{Ew_isospin_N4}
\end{figure}

\begin{figure}[htb]
\centering
\subfigure[$K=0$\, ($\omega=0$)]{\includegraphics[totalheight=3cm]{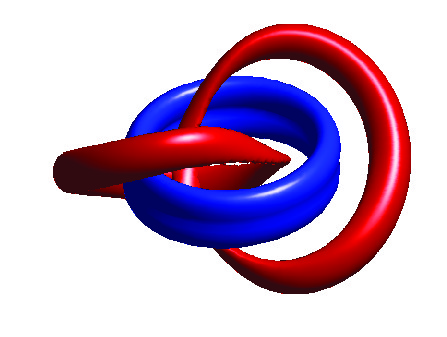}}
\subfigure[$K=10$\, ($\omega=0.292$)]{\includegraphics[totalheight=3cm]{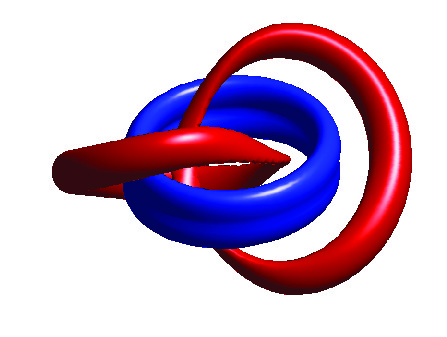}}
\subfigure[$K=20$\, ($\omega=0.554$)]{\includegraphics[totalheight=3cm]{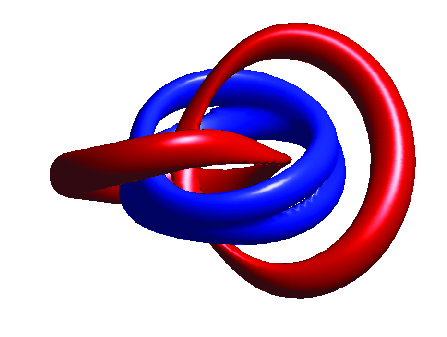}}
\subfigure[$K=21$\, ($\omega=0.572$)]{\includegraphics[totalheight=3cm]{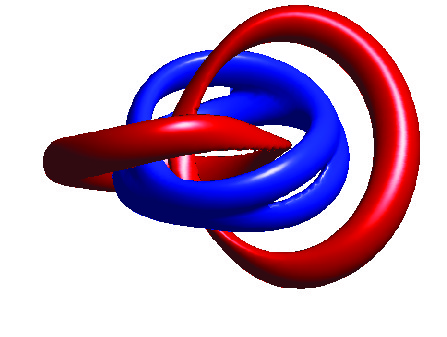}}\\
\subfigure[$K=22$\, ($\omega=0.586$)]{\includegraphics[totalheight=3cm]{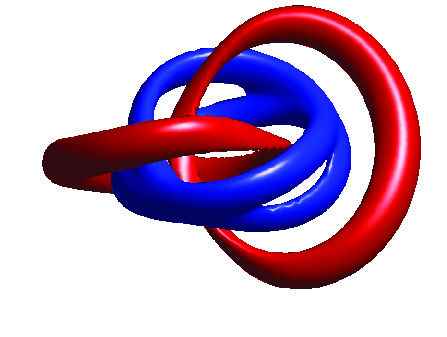}}
\subfigure[$K=25$\, ($\omega=0.609$)]{\includegraphics[totalheight=3cm]{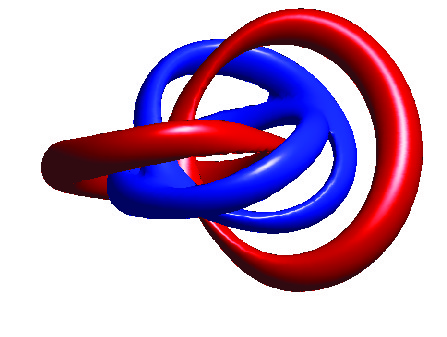}}
\subfigure[$K=30$\, ($\omega=0.654$)]{\includegraphics[totalheight=3cm]{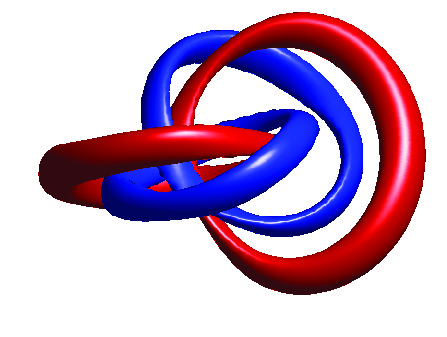}}
\subfigure[$K=40$\, ($\omega=0.755$)]{\includegraphics[totalheight=3cm]{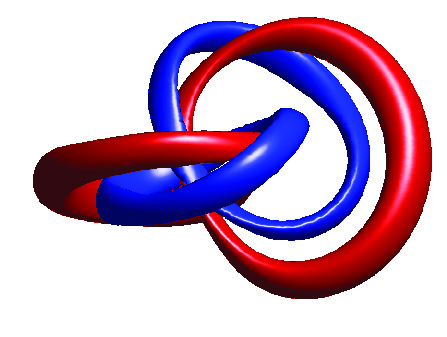}}
\caption{Deformation of the isospinning $4\mathcal{A}_{2,2}$ Hopf configuration into $4\mathcal{L}^{1,1}_{1,1}$. Results are plotted for potential $V_I$ (results for potential $V_{II}$ show the same qualitative behaviour). We visualize the linking structure by plotting tubelike isosurfaces $\phi_1=-0.9$ (red tube) and $\phi_3=-0.9$ (blue tube). Recall that the angular momentum $K$ is given in units of $4\pi$.}
\label{Iso_isospin_A22}
\end{figure}

\begin{figure}[htb]
\centering
\subfigure[$E_\text{{tot}}$ as a function of $\omega$]{\includegraphics[totalheight=6.cm]{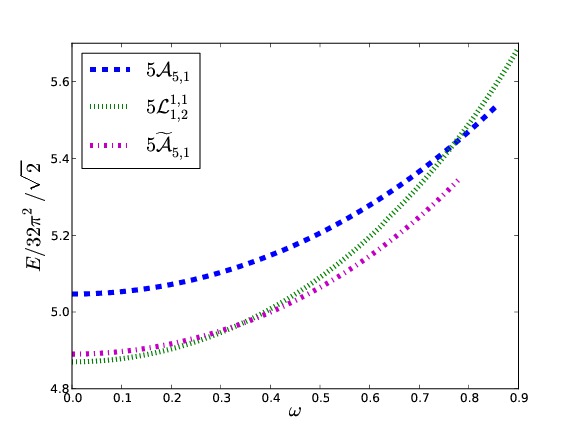}}
\subfigure[$E_\text{{tot}}$ as a function of $K$]{\includegraphics[totalheight=6.cm]{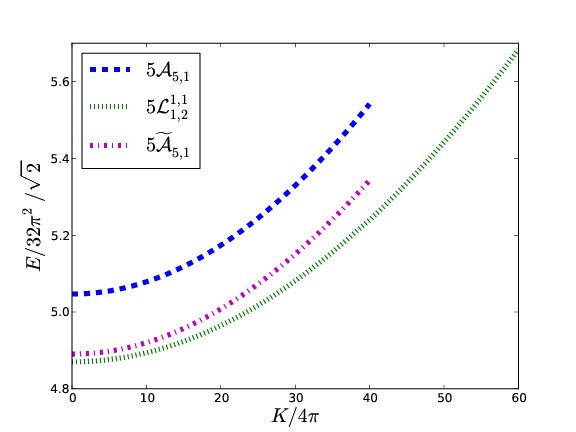}}
\caption{Total energy $E_\text{tot}$ of isospinning charge 5-Hopf Solitons as a function of $\omega$ and $K$ ($V=V_{I}$). The energy curve $E_\text{tot}(\omega)$ of the link $5\mathcal{L}^{1,1}_{1,2}$ (green curve) crosses the one of $5\mathcal{\widetilde{A}}_{5,1}$ (purple curve) at $\omega\approx0.33$. The lowest-energy, isospinning soliton is of type $5\mathcal{L}^{1,1}_{1,2}$ for $\omega\in[0,0.33)$ and $5\mathcal{\widetilde{A}}_{5,1}$ for $\omega\in[0.33,1)$. For comparison, we also show the axial, unstable $5\mathcal{A}_{5,1}$ solution (blue curve).}
\label{Ew_isospin_N5}
\end{figure}

\begin{figure}[htb]
\centering
\setcounter{subfigure}{0}
\subfigure[$E_\text{tot}$ as a function of $\omega$]{\includegraphics[totalheight=6cm]{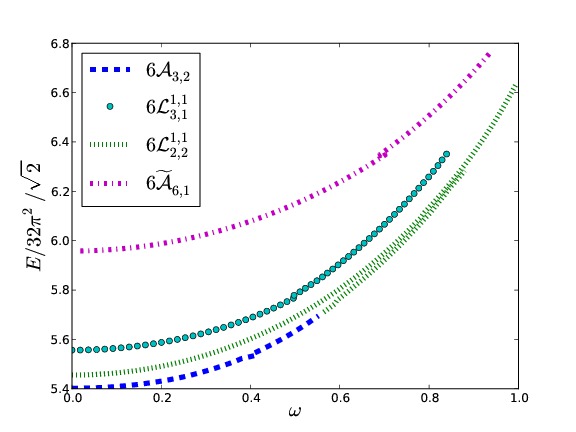}}
\subfigure[$E_\text{tot}$ as a function of $K$]{\includegraphics[totalheight=6cm]{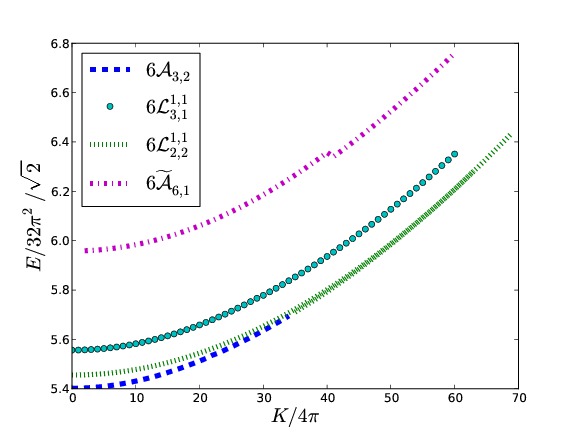}}
\caption{Total energy $E_\text{tot}$ for different, isospinning 6-Hopf soliton configurations ($V=V_I$). The isospinning solution of type $6\mathcal{A}_{3,2}$ (blue curve) deforms into $6\mathcal{L}^{1,1}_{2,2}$ (green curve) at $\omega\approx0.56$ ($K\approx35$). The $\mathcal{\widetilde{A}}_{6,1}$ configuration (purple curve) exists for all $\omega\in[0,1)$, but is of higher energy.}
\label{Ew_isospin_N6}
\end{figure}

\begin{figure}[htb]
\centering

\subfigure{\includegraphics[totalheight=3.0cm]{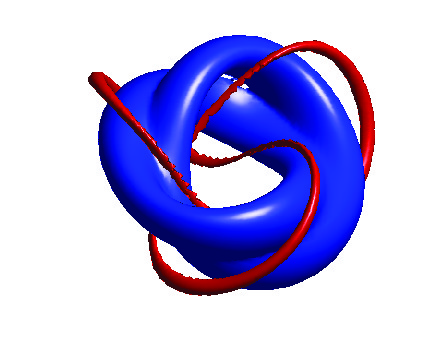}}
\subfigure{\includegraphics[totalheight=3.0cm]{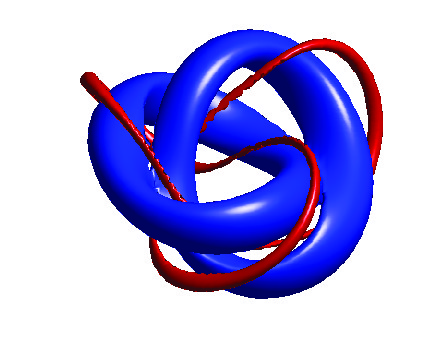}}
\subfigure{\includegraphics[totalheight=3.0cm]{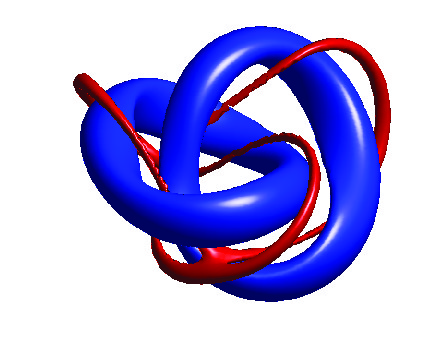}}
\subfigure{\includegraphics[totalheight=3.0cm]{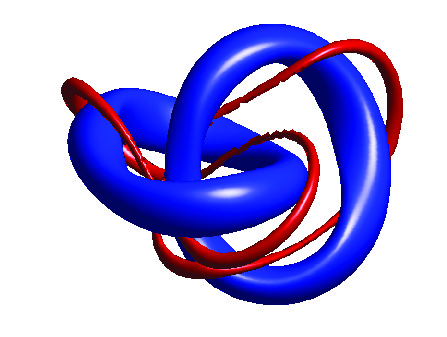}}\\
\setcounter{subfigure}{0}
\subfigure[$K=0$\, ($\omega=0$)]{\includegraphics[totalheight=3.0cm]{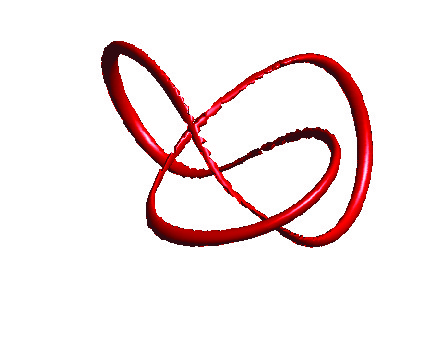}}
\subfigure[$K=20$\, ($\omega=0.367$)]{\includegraphics[totalheight=3.0cm]{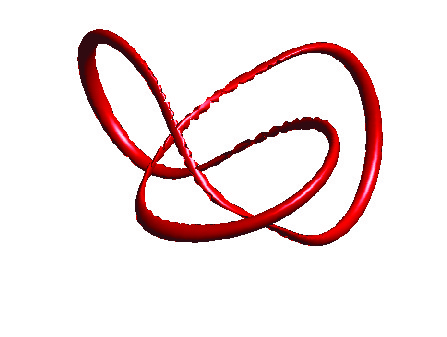}}
\subfigure[$K=30$\, ($\omega=0.502$)]{\includegraphics[totalheight=3.0cm]{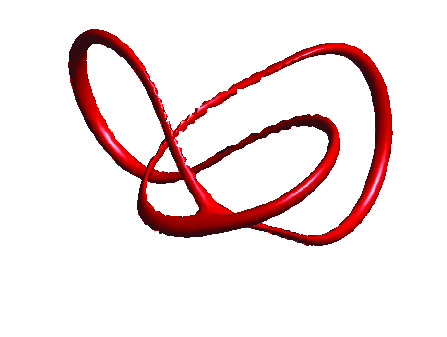}}
\subfigure[$K=45$\, ($\omega=0.677$)]{\includegraphics[totalheight=3.0cm]{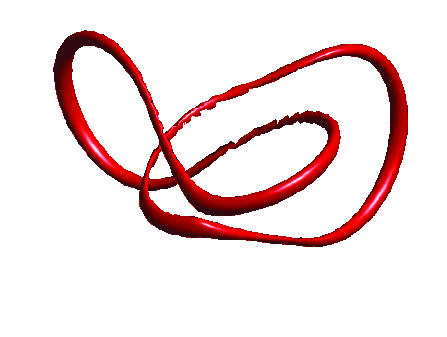}}
\caption{Deformation of the isospinning  $6\mathcal{A}_{3,2}$ Hopf soliton solution into $6\mathcal{L}^{1,1}_{2,2}$. First row: We display isosurfaces $\phi_1=-0.98$ (red tube) and $\phi_3=-0.65$ (blue tube) to illustrate the change of the solution types. Second row: We show the linking curves for $\phi_1=-0.98$, separately.   }
\label{N6_def}
\end{figure}

\begin{figure}[htb]
\centering
\subfigure[$K=0$\, ($\omega=0$)]{\includegraphics[totalheight=3.0cm]{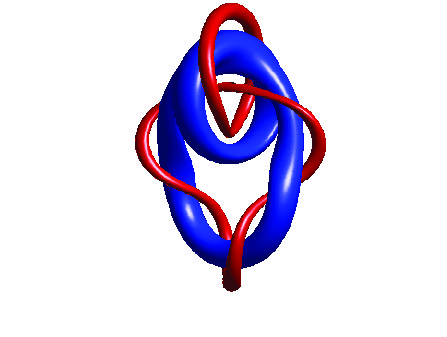}}
\subfigure[$K=40$\, ($\omega=0.619$)]{\includegraphics[totalheight=3.0cm]{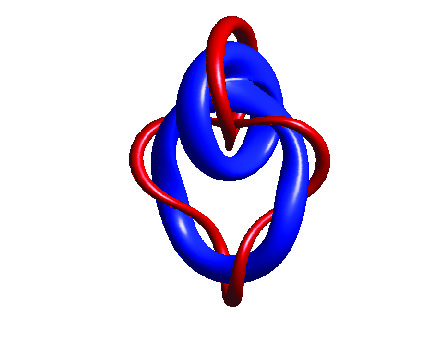}}
\subfigure[$K=45$\, ($\omega=0.677$)]{\includegraphics[totalheight=3.0cm]{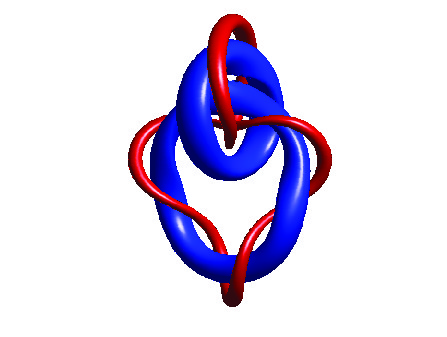}}
\subfigure[$K=59$\, ($\omega=0.829$)]{\includegraphics[totalheight=3.0cm]{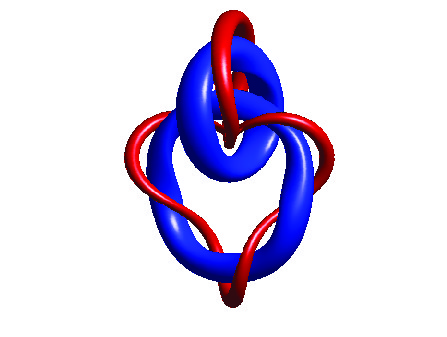}}
\caption{Linking curves of the isospinning  $6\mathcal{L}^{1,1}_{3,1}$ Hopf soliton solution. The linking structure is visualized by the isosurfaces $\phi_1=-0.95$ (red tube) and $\phi_3=-0.65$ (blue tube).}
\label{N6_Links_def}
\end{figure}


\subsection{Higher Charge Hopf Solitons: $4\leq N\leq 8$}

\begin{figure}[htb]
\centering
\subfigure[$E_\text{{tot}}$ as a function of $\omega$]{\includegraphics[totalheight=6.cm]{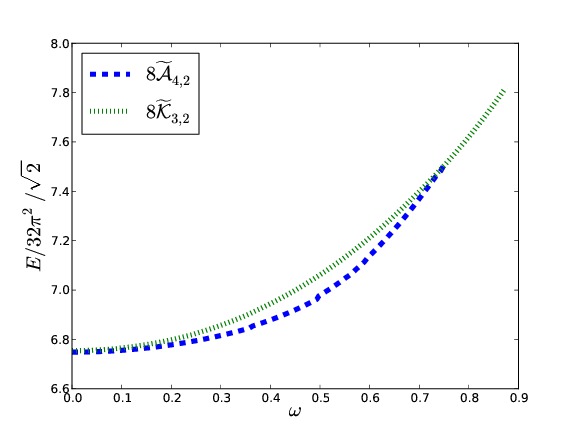}}
\subfigure[$E_\text{{tot}}$ as a function of $K$]{\includegraphics[totalheight=6.cm]{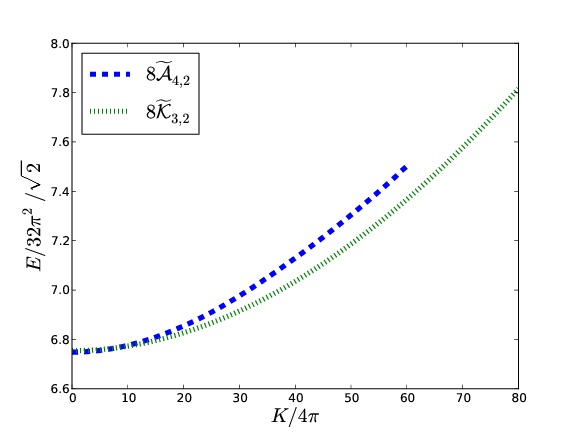}}
\caption{The total energy $E_{\text{tot}}$ for isospinning 8-Hopf solitons as function of $\omega$ and $K$ ($V=V_{I}$). The isospinning $8\mathcal{\widetilde{A}}_{4,2}$ configurations deforms into the $8\mathcal{K}_{3,2}$ knot at $\omega\approx0.68$ ($K\approx 49$).  }
\label{Ew_isospin_N8}
\end{figure}

\begin{figure}[htb]
\centering
\subfigure[$K=0$\, ($\omega=0$)]{\includegraphics[totalheight=3.0cm]{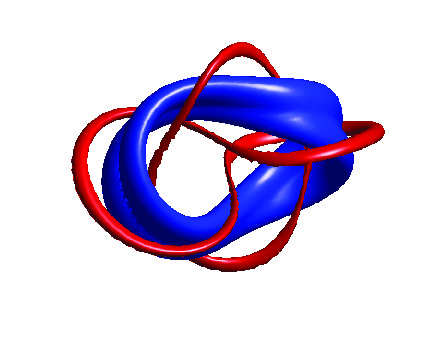}}
\subfigure[$K=30$\, ($\omega=0.504$)]{\includegraphics[totalheight=3.0cm]{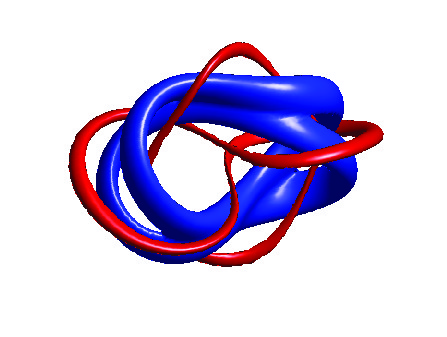}}
\subfigure[$K=40$\, ($\omega=0.615$)]{\includegraphics[totalheight=3.0cm]{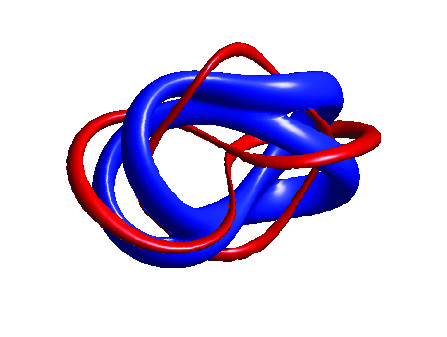}}
\subfigure[$K=60$\, ($\omega=0.780$)]{\includegraphics[totalheight=3.0cm]{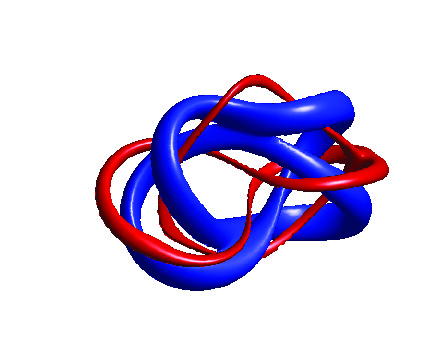}}
\caption{Deformation of the isospinning $8\mathcal{\widetilde{A}}_{4,2}$ Hopf configuration into $8\mathcal{K}_{3,2}$. The linking structure is visualized by the isosurfaces $\phi_1=-0.97$ (red tube) and $\phi_3=-0.8$ (blue tube). }
\label{N8_def}
\end{figure}

\begin{itemize}
\item $N=4$: The energy and moment of inertia plots for isospinning 4-Hopf solitons ($4\mathcal{\widetilde{A}}_{4,1},\,4\mathcal{A}_{2,2}$) are shown in Fig~\ref{Ew_isospin_N4}. The $4\mathcal{\widetilde{A}}_{4,1}$ configuration is found to be the solution type of lowest energy for all $\omega$ and $K$. The $4\mathcal{A}_{2,2}$ soliton deforms for $\omega\geq 0.60$ ($K\geq 23$) into a $4\mathcal{L}^{1,1}_{1,1}$ link, which means into a solution type which does not represent a local minimum in the static case ($\omega=0$). The isosurface plots in Fig.~\ref{Iso_isospin_A22} illustrate the formation of the linked configuration as $K$ increases.

\item $N=5$: We show in Fig.~\ref{Ew_isospin_N5} the total energy $E_{\text{tot}}$ of isospinning charge-5 Hopf solitons ($5\mathcal{L}^{1,1}_{1,2},\,5\mathcal{\widetilde{A}}_{5,1},\,5\mathcal{A}_{5,1}$) as a function of the rotation frequency $\omega$ and the angular momentum $K$. We observe that the energy curve $E_\text{tot}(\omega)$ of the linked unknot $5\mathcal{L}^{1,1}_{1,2}$ crosses the one of the bent ring $\mathcal{\widetilde{A}}_{5,1}$ at $\omega\approx0.33$. For $\omega>0.33$ the bent ring becomes the new ground state for Hopf charge $N=5$. However, for fixed $K$ the linked configuration continues to be the lowest energy state, see Fig.~\ref{Ew_isospin_N5}.

\item $N=6$: Our simulations of isospinning 6-Hopf solitons ($6\mathcal{A}_{3,2},\,6\mathcal{L}^{1,1}_{3,1},\,6\mathcal{L}^{1,1}_{2,2},\,6\mathcal{\widetilde{A}}_{6,1}$) are summarized by the energy curves in Fig.~\ref{Ew_isospin_N6}. Here, we see an example of transmutation: the $6\mathcal{A}_{3,2}$ configuration that is the ground state at $\omega=0$ transforms into a $6\mathcal{L}^{1,1}_{2,2}$ link when K increases. For $K\geq35$ ($\omega\geq0.56$) the $6\mathcal{A}_{3,2}$ soliton has completely deformed into the link configuration that forms the new lowest energy state. The deformation process is visualized by the isosurface plots in Fig.~\ref{N6_def}. Bent Hopf configurations of solution type $6\mathcal{\widetilde{A}}_{6,1}$ and links of type $6\mathcal{L}^{1,1}_{3,1}$  have higher energies for all $\omega$ and $K$. The linking curves in Fig.~\ref{N6_Links_def} show that the $6\mathcal{L}^{1,1}_{3,1}$ configuration is of the same qualitative shape for all $\omega$ and $K$.  
\item $N=7$: We do not observe any crossing of the energy curves of isospinning $7\mathcal{K}_{3,2}$ and $7\mathcal{K}_{2,3}$ knot solutions. We find the $7\mathcal{K}_{3,2}$ knot as the state of lowest energy for all $\omega$ and $K$.  

\item $N=8$: We display in Fig.~\ref{Ew_isospin_N8} the total energies of isospinning $8\mathcal{\widetilde{A}}_{4,2}$ and $8\mathcal{K}_{3,2}$ configurations as a function of angular frequency and momentum. At $\omega=0$ the configurations can be seen as energy-degenerate\footnote{By this we mean that they have the same energy within the numerical errors and hence  we are unable to distinguish them.}, but in the isospinning case we find that the $8\mathcal{K}_{3,2}$ solution type has a higher energy than the $8\mathcal{\widetilde{A}}_{4,2}$ configuration. In fact, we can see that the $8\mathcal{\widetilde{A}}_{4,2}$ solution slowly deforms into the  knotted solution, as illustrated in Fig.~\ref{N8_def} by plotting the linking structure. The transition occurs at $\omega\approx0.68$ ($K\approx49$).  
\end{itemize}

\section{Conclusions}\label{Sec_Con}
We have performed full three-dimensional numerical relaxations of isospinning soliton solutions in the Skyrme-Faddeev model with mass terms included. Our computations of charge-4, -6 and -8  solitons show that the qualitative shapes of internally rotating Hopf solitons can differ from the static ($\omega=0$) solitons. However, in most cases (for Hopf charges $N=1,2,3,5,7$) the solution types present at $\omega=0$ also exist for non-zero $\omega$. The qualitative shape of the lowest energy configuration can be frequency dependent. The energy curves $E_\text{tot}(\omega)$ for a given $N$ can cross and minima can swap (e.g., $N=5$). In summary, we distinguish three different types of behavior:
\begin{itemize}
\item \emph{Crossings of $E_{\text{tot}}(\omega)$:} The energy curves $E_{\text{tot}}(\omega)$ of Hopf solitons for different solution types of the same charge $N$ can cross, which results in a rearrangement of the spectrum of minimal-energy configurations. Our simulations on isospinning charge-5 solitons illustrate this: at $\omega=0$ the link $5\mathcal{L}^{1,1}_{1,2}$ is the lowest energy solution, but for $\omega\approx0.33$ its energy curve crosses that of the bent unknot $5\mathcal{\widetilde{A}}_{5,1}$. For $\omega\geq0.33$ the lowest energy soliton is given by $5\mathcal{\widetilde{A}}_{5,1}$.
\item \emph{Transmutation:} Isospinning Hopf solitons can deform into minimal-energy solutions of a type that also exists at $\omega=0$  (e.g. $6\mathcal{A}_{3,2}\rightarrow6\mathcal{L}^{1,1}_{2,2},\,8\mathcal{\widetilde{A}}_{4,2}\rightarrow8\mathcal{K}_{3,2}$). 
\item \emph{Formation of new solution types:} New solution types can emerge which are unstable for vanishing $\omega$. For example, for $N=4$ the $4\mathcal{A}_{2,2}$ deforms into $4\mathcal{L}^{1,1}_{1,1}$ with the later only being stable for $\omega\geq0.60$ 
\end{itemize}

Naturally one expects these effects to be present and increasingly relevant for higher Hopf charges ($N>8$) since the number of (local) energy minima grows with the Hopf charge $N$ \cite{Sutcliffe:2007ui}. 

In this article we have focussed on purely classically isospinning soliton solutions in the Skyrme-Faddeev model. The relevance of classically (iso)spinning soliton solutions was discussed in Ref.~\cite{Manton:2011mi} in the context of the Skyrme model. There it was argued that classically spinning Skyrmions could be used to model classically the quantized Skyrmion states. For example, a spin-1/2 proton in its spin up state can be interpreted within this approximate classical description as a hedgehog Skyrmion of topological charge $B=1$ spinning anticlockwise relative to the positive $z$ axis and with its normalized pion fields $\boldsymbol{\hat{\pi}}$ orientated in such a way that $\boldsymbol{\pi}_3=\pm1$ for $z\rightarrow\pm\infty$, respectively. Analogously,  classically spinning Hopf soliton solutions can classically model the quantized spectra of glueballs \cite{Cho:1979nv,Cho:1980nx,Cho:1981ww,Shabanov:1999uv,Langmann:1999nn,Faddeev:2001dda,Cho:2004ue,Kondo:2004dg}. To do this, it is necessary to determine the (iso)space orientations that describe the excited states of glueballs. To approximate states of non-vanishing spin, rotations in physical space have to be implemented in our computations, which significantly complicates our numerics.

Our numerical results are of relevance for the quantization of the classical soliton solutions. There are two main methods used in the literature to obtain quantized Hopf solitons: the bosonic, semiclassical collective coordinate quantization \cite{Su:2001zw,Kondo:2006sa} and the fermionic quantization \cite{Krusch:2005bn} that is based on the Finkelstein-Rubinstein (FR) approach \cite{Finkelstein:1968hy}. Both approaches assume that the symmetries of the classical Hopf configurations are not broken by centrifugal effects. In the semiclassical bosonic collective coordinate quantization procedure glueballs can be modeled by quantum mechanical states on the moduli space -- the finite-dimensional space of static minimal energy Hopf solutions in a given topological sector which is generated from a single Hopf configuration by rotations and isorotations. The effective Hamiltonian on this restricted configuration space is canonically quantized. The numerical calculations presented in this paper could be seen as a classical approximation to the collective coordinate dynamics on the moduli space. The allowed quantum states have to satisfy the FR constraints \cite{Finkelstein:1968hy} which follow from the continuous and discrete symmetries of the classical Hopf configurations: For a bosonic quantum theory the FR constraints result in constraints for the wave functions defined on the configuration space, whereas fermionic quantization \cite{Krusch:2005bn} constrains the wave functions on the covering space of configuration space.

Ground states and first excited states of Hopf solitons for charges up to $7$ have been calculated in Ref.~\cite{Krusch:2005bn} using the symmetries of the classical Hopf solutions given in Ref.~\cite{Hietarinta:2000ci}. It would be instructive to work out the spectra that emerge from the classical solutions calculated in our article. The isospinning, minimal energy Hopf solitons of charge $N=5,6,8$ are particularly interesting since their symmetries are different from those of the static configurations which are commonly used to calculate the solitons' possible ground states. The presentation of a self-consistent, non-rigid quantization procedure goes far beyond the scope of this paper and is the subject of future research.

\section*{Note added}
Similar results were also reported in a very recent paper \cite{Harland:2013uk} which appeared when our paper was in preparation. The authors in \cite{Harland:2013uk} carried out most of their calculations with $\mu=2$ and the potential choice $V_{II}$. Differences to our results are that they neither identify a $6\mathcal{A}_{3,2}$ nor  a $8\mathcal{\widetilde{A}}_{4,2}$ configuration. Unfortunately they did not visualize the linking structure of their 6- and 8-Hopf soliton solutions, so that we could not compare them. Differences to our results could be due to the different potential choice or to the different choice of the mass parameter $\mu$.

\section*{Acknowledgements}
We would like to acknowledge the use of the National Supercomputing Centre in Cambridge. We thank Juha J\"aykk\"a, Yakov Shnir and Paul Sutcliffe for useful discussions.

\end{document}